\documentclass{iaus}
\usepackage{graphicx}

\title[~~Wide-field spectroscopic surveys] 
{Panchromatic properties of galaxies in wide-field optical \\ spectroscopic and photometric surveys}

\author[Simon P. Driver]   
{Simon P. Driver$^{1,2}$}

\affiliation{$^1$ International Centre for Radio Astronomy Research (ICRAR), University of Western Australia, 35 Stirling Highway, Crawley, Perth, Australia, WA 6009 \\ email: {Simon.Driver@icrar.org}
\\[\affilskip]
$^2$ School of Physics and Astronomy, University of St Andrews, North Haugh, St Andrews, UK, KY16 8RS, email: {spd3@st-and.ac.uk}}

\pubyear{2011} 
\volume{284}  
\pagerange{1--12}
 \jname{The Spectral Energy Distribution of
  Galaxies} \editors{R.J. Tuffs \& C.C.Popescu, eds.}
\begin{document}

\maketitle

\begin{abstract}
The past 15 years have seen an explosion in the number of redshifts
recovered via wide area spectroscopic surveys. At the current time
there are approximately 2million spectroscopic galaxy redshifts known
(and rising) which represents an extraordinary growth since the
pioneering work of Marc Davis and John Huchra. Similarly there has
been a parallel explosion in wavelength coverage with imaging surveys
progressing from single band, to multi-band, to truly multiwavelength
or pan-chromatic involving the coordination of multiple
facilities. With these empirically motivated studies has come a wealth
of new discoveries impacting almost all areas of astrophysics. Today
individual surveys, as best demonstrated by the Sloan Digital Sky
Survey, now rank shoulder-to-shoulder alongside major facilities. In
the coming years this trend is set to continue as we being the process
of designing and conducting the next generation of spectroscopic
surveys supported by multi-facility wavelength coverage.

\end{abstract}

\firstsection 

\section{Introduction}
This article briefly summarises the development of wide field
spectroscopic survey programs in extragalactic astronomy (see Fig.~1,
Table.~1), and their impact as measured in publications and citations
(see Table.~2). The statistics shown in Table.~2 are necessarily crude
but indicative of the major revolution which is overtaking our subject
--- the rise of surveys over facilities. Whereas once the key question
senior astronomers might ask of each other seemed to be ``so what are
you building?'', in this age of internationally funded
mega-facilities, the more pertinent question might be ``so what survey
are you designing?''. The likely answer will be one which combines
both spectroscopic information with panchromatic imaging.
Traditionally imaging surveys, as pioneered by the various Schmidt
surveys, were single bandpass (${b_J}$) and then later double
bandpass ($b_J,r_F$). These were followed by a move to
multi-band programs (e.g., $BVRI$, $ugriz$, $YJHK$) which used multiple
filters to fill in the wavelength gaps but still remain single
facility imaging surveys. Truly multiwavelength surveys span more than
one imaging facility, and panchromatic surveys span multiple
facilities covering a significant fraction of the em-spectrum (e.g.,
GOODS; Giavalisco et al.,~2004; GAMA; Driver et al.,~2011). This
trend, towards broader spectral coverage, is driven not only by
technological advancement but also by an appreciation that galaxies in
particular emit significant levels of radiation at almost all
wavelengths and to truly understand their nature will require robust
distances combined with total energy measurements.

\begin{figure}[h]

\vspace*{-3.5 cm}

\includegraphics[width=5.0in]{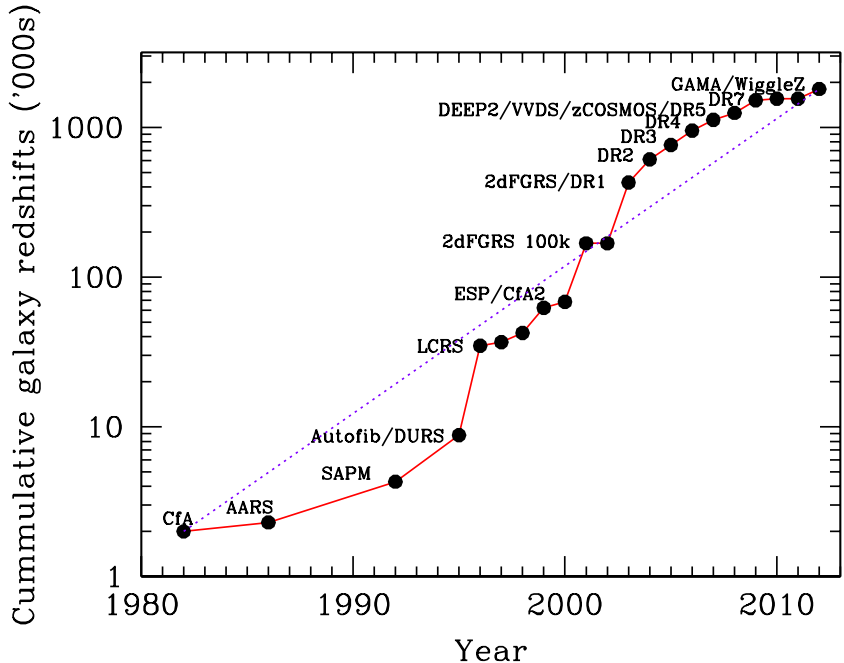} 

 \caption{The rise in galaxy redshifts over the past 30 years.}
   \label{fig1}

\end{figure}

\section{Do we still need spectroscopic surveys?}
Figure~1 shows the increase in the cumulative number of known galaxy
redshifts (AGN are not included here) and the prominent surveys are
listed in Table~1. The most notable are of course the Sloan Digital
Sky Survey (SDSS; York et al.,~2000), and the two-degree field galaxy
redshift survey (2dFGRS; Colless et al.,~2001). Between them these
surveys account for over half the known redshifts and even more
importantly both surveys have very clean selection criteria ($r_{AB} <
17.77$ and $b_{J} < 19.6$ respectively).

\begin{table}[h]
  \begin{center}
  \caption{Major spectroscopic surveys motivated from UV/optical or near-IR imaging data}
  \label{tab1}
 {\scriptsize
  \begin{tabular}{|l|l|c|c|r|}\hline 
{\bf Survey} & {\bf Reference} & {\bf Facility} & {\bf Selection} & {\bf Redshifts} \\ \hline
CfA & Davis et al.,~(1982) & Mt Hopkins 1.5m/Z-machine & $B<14.5$ & 2k \\
AARS & Peterson et al.,~(1986) & AAT/RGO & $B_J < 17.0$ & 0.3k \\
SARS & Loveday et al.,~(1992) & MSSSO 2.3m/DBS & $b_J < 17.15$ & 2k \\
Autofib & Ellis et al.,~(1996) & AAT/Autofib & $b_J < 22.0$ & 1k  \\
DURS & Ratcliffe et al.,~(1996) & UKST/FLAIR & $b_J < 17.0$ & 2.5k \\
LCRS & Schectman et al.,~(1996) & DuPont/MOS & $R<17.5$ & 26k \\ 
CFRS & Lilly et al.,~(1996) & CFHT/MOS & $I_{AB} < 22.5$ & 1k\\ 
CS & Geller et al.,~(1997) & Various & $r<16.13$ & 2k \\
ESP & Vettolani et al.,~(1997) &  ESO 3.6m/OPTOPUS & $b_J < 19.4$ & 4k \\
SSRS2 & da Costa et al.,~(1998) & Various & $B<15.5$ & 5.5k \\
CfA2 & Falco et al.,~(1999) & Various & $B < 15.5$ & 20k \\
CNOC2 & Yee et al.,~(2000) & CFHT/MOS & $R < 21.5$ & 6k \\
2dFGRS & Colless et al.,~(2001) & AAT/2dF & $b_J < 19.6$ & 227k \\
SDSS Main & Strauss et al.,~(2002) & SDSS 2.5m/Spectrographs & $r<17.77$ & 930k \\
SDSS LRG & Eisenstein et al.,~(2002) & SDSS 2.5m/Spectrographs & $r < 19.5$ & 120k \\
DEEP1\&2 & Davis et al.,~(2003) & Keck/Deimos & $R_{AB}<24.1$ & 51k \\
H-AAO & Huang et al.,~(2003) & AAT/2dF & $K<15.0$ & 1k  \\
MGC & Driver et al.,~(2005) & AAT/2dF & $B<20.0$ & 10k \\
SDSS Stripe82 & Baldry et al.,~(2005) & SDSS/Spectrographs & $u < 20.5$ & 70k \\
2SLAQ-LRG & Cannon et al.,~(2006) & AAT/2dF & $i<19.8$ & 13k \\
6dfGRS & Jones et al.,(2009) & UKST/6dF & $K < 12.75$+ & 110k \\
VVDS-wide & Garilli et al.,(2008) & VLT/VIMOS & $I_{AB} < 22.5$ & 35k \\
VVDS-deep & Le Fevre et al.,(2005) & VLT/VIMOS & $I_{AB} < 24.0$ & 12k(150k) \\
VVDS-ultradeep & Le Fevre et al.,(2005) & VLT/VIMOS & $I_{AB} < 24.75$ & 0k(1k) \\
zCOSMOS-bright & Lilly et al.,~(2007) & VLT/VIMOS & $I_{AB} < 22.5$ & 10k(20k) \\
zCOSMOS-deep & Lilly et al.,~(2007) & VLT/VIMOS & $I_{AB} < 24.0$ & 1k(10k) \\
WiggleZ & Drinkwater et al. (2010) & AAT/AAOmega & $r<22.5$+ & 250k \\ 
AGES & Kochaneck et al. (2011) & MMT/Hectospec & $R <20$+ & 19k \\
GAMA & Driver et al. (2011) & AAT/AAOmega & $r<19.8$ & 180k(350k) \\ 
Vipers & N/A & VLT/VIMOS & $I_{AB} < 22.5$ & 20k(100k) \\
BOSS & N/A & SDSS 2.5m/Spectrographs & $i<20$ & 500k(1500k) \\ \hline
  \end{tabular}
  }
 \end{center}
\vspace{1mm}
\noindent
The majority of the information shown in this table was kindly
provided by Ivan Baldry (see
http://www.astro.ljmu.ac.uk/$\sim$ikb/research/galaxy-redshift-surveys.html)
which was used to produce Fig.~1 in Baldry et al.,~(2010).
\end{table}

However this rise in spectroscopic surveys has been matched by an
expansion in wavelength coverage which has led to the realisation of
photometric redshifts (e.g., ANNz; Collister \& Lahav 2004, see Fig.~2
(left)). Photometric redshifts allow one to bypass the laborious
process of conducting a spectroscopic program by using the continuum
shape (and in particular the 4000\AA~break and the Lyman-limit) to
estimate distance. Comparisons between photometric and spectroscopic
surveys are impressive with typical accuracies running at $\Delta
z/(1+z) \pm 0.03$ (Collister \& Lahav 2004). Purpose built photo-z
surveys such as COMBO-17 (Wolf et al 2003), which uses a 17 medium
band filter set --- essentially producing a very low dispersion
spectrum --- can improve upon this accuracy even further ($\Delta
z/(1+z) \sim 0.01$; Wolfe et al., 2008). This rise in photometric
redshifts begs the question as to whether we still actually need large
scale spectroscopic programs rather than simply conducting spot test
checks of photo-z cats. For large scale statistical studies this is
probably true but the following three points argue for some caution:

\noindent
{\bf 1) At low redshift:} At low redshift the typically photo-z error
$\Delta z/(1+z) \sim \pm 0.03$ becomes significant. This is easily
demonstrated by constructing a luminosity function using entirely
photometric or spectroscopic redshifts via a simple $1/V_{\rm max}$
estimator.  Fig.~2 (left) shows the photo-z v spec-z comparison for
53k galaxies from the GAMA survey which generally appears well
behaved. However Fig.~2 (right) shows the corresponding derived
luminosity functions with somewhat disastrous implications for the
faint-end slope. This is because photo-z's are generally calibrated
for the most numerous galaxy type within ones' training set. Galaxies
with divergent properties can have significant systematic errors
leading to the kind of severe bias shown in Fig.~2 (right).

\begin{figure}[h]

\vspace*{-6.0 cm}

\includegraphics[width=5.0in]{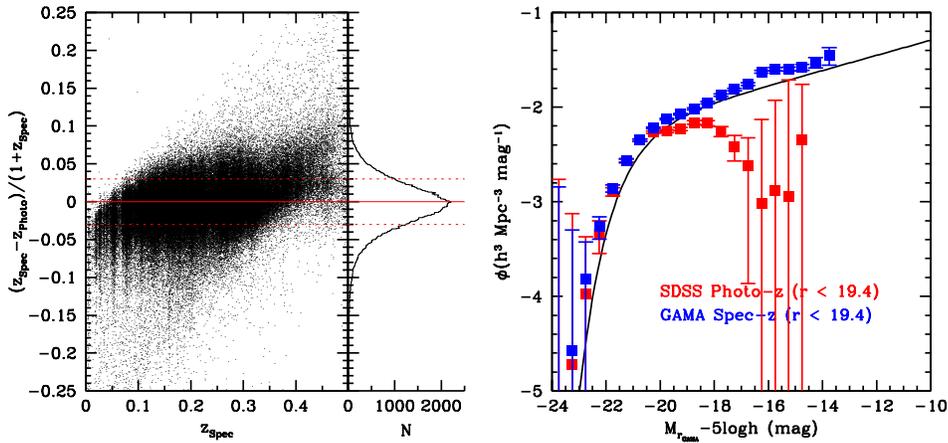} 

 \caption{(left) Photometric versus Spectroscopic redshifts for
   170,000 galaxies drawn from the GAMA database. The accuracy is $\pm
   0.03$. (right) the resulting luminosity functions derived for the
   two samples indicating good agreement at $L^*$ but dramatically
   different faint-ends.}
   \label{fig2}

\end{figure}

\noindent
{\bf 2) When fidelity is required:} Fig.~3 shows a typical cone
diagram for the GAMA 12hr region using photometric (upper) or
spectroscopic (lower) redshifts. While statistically one can still
recover some sense of the generic clustering properties (suitable for
cosmology) it is clearly impractical to use photo-z's to identify
individual filaments, clusters, or groups and their associated
masses. High fidelity studies are likely to be a key focus area in the
coming years as we look to distinguish between the influence of
environment and halo mass (Haas, Schaye \& Jeeson-Daniel 2011).

\begin{figure}[h]

\vspace*{-2.0 cm}

\includegraphics[width=5.0in,angle=-90]{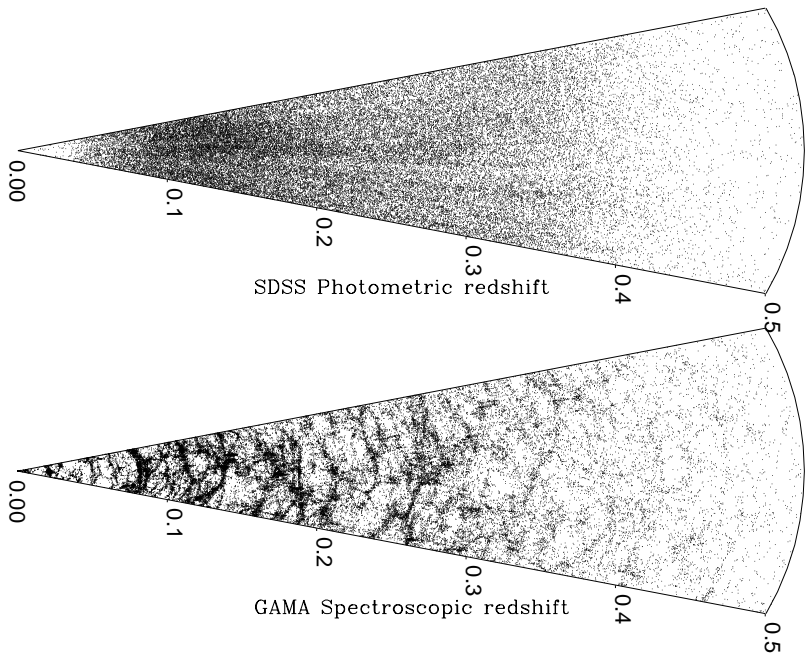} 

\vspace*{-2.0 cm}

 \caption{A comparison of photometric redshifts (upper) and spectroscopic redshifts (lower) derived from the SDSS and GAMA surveys.}
   \label{fig3}

\end{figure}

\noindent
{\bf 3) When the test galaxies extend beyond the calibration data.}
Photometric redshifts make use of the 4000\AA~break and are typically
calibrated to bright and local spectroscopic redshift samples which
exhibit strong 4000\AA~breaks. However the break is less apparent in
star-forming systems. As one progresses either to lower mass systems
in the nearby Universe or to high redshift massive systems one moves
to more star-forming populations which exhibit flatter spectra and the
likelihood of confusion increases.

\section{High impact photo-z and spectro-z surveys}
Despite these concerns photo-z's have proved invaluable in many areas
and in particular for the interpretation of data seen in the very deep
fields obtained by the Hubble Space Telescope. Table~2 attempts to
provide some indication of the impact of many of the spectroscopic and
photometric redshift surveys over the past few decades. The landscape
is generally dominated by two nearby spectroscopic programmes (SDSS,
2dFGRS), coupled with deep HST studies (HDF, GOODS, COSMOS), and
forays into new wavelengths (2MASS, GOODS, GALEX) all of which build
upon the pioneering work on the CfA surveys.

\begin{table}[h]
  \begin{center}
  \caption{Major extragalactic optical/near-IR surveys and their impact in terms of refereed papers and citations}
  \label{tab1}
 {\scriptsize
  \begin{tabular}{|l|r|r|}\hline 
{\bf Survey} & {\bf Papers} & {\bf Citations} \\ \hline
Sloan Digital Sky Survey (SDSS) & 3254 & 135330 \\
Hubble Deep Field (HDF) & 590 & 46360 \\
Two Micron All Sky Survey (2MASS) & 968 & 27253 \\
Two degree Field Galaxy Redshift Survey (2dFGRS) & 197 & 23219 \\
Great Observatories Origins Deep Survey (GOODS) & 498 & 24578 \\
Centre for Astrophysics (CfA) & 263 & 11989 \\
Galaxy Evolution Explorer GALEX) & 465 & 11560 \\
Canada France Redshift Survey (CFRS) & 90 & 8493 \\
Cosmic Evolution Survey (COSMOS) & 241 & 7354 \\
Ultra Deep Field (UDF) & 188 & 7237 \\
Classifying Objects by Medium-Band Observations (COMBO-17) & 86 & 6088 \\
Deep Evolutionary Exploratory Probe (DEEP) & 113 & 5118 \\
Las Campanas Redshift Survey (LCRS) & 96 & 4063 \\
APM Galaxy Survey (APMGS) & 57 & 4058 \\
VIMOS VLT Deep Survey (VVDS) & 75 & 3689 \\
UKIRT Infrared Deep Sky Survey (UKIDSS) & 105 & 3444 \\
Southern Sky Redshift Survey 2 (SSRS2) & 72 & 3421 \\
Point Source Catalogue Survey (PSCz) & 94 & 3270 \\
2dF QSO redshift Survey (2QZ) & 53 & 2930 \\
Galaxy Evolution from Morphologies and SEDs (GEMs) & 62 & 2696 \\
NOAO Deep Wide Field Survey (NDFWS) & 77 & 2555 \\
CFHT Legacy Survey (CFHTLS) & 72 & 2442 \\
Millennium Galaxy Catalogue (MGC) & 32 & 1590 \\
zCOSMOS & 47 & 1567 \\
Stromlo-APM Redshift Survey (SARS) & 19 & 1389 \\
AGN and Galaxy Evolution Survey (AGES) & 31 & 1337 \\ 
Gemini Deep Deep Survey (GDDS) & 20 & 1304 \\
The 2dF-SDSS LRG And QSO Survey (2SLAQ) & 33 & 1280 \\
Canadian Network for Observational Cosmology Field Galaxy Redshift Survey (CNOC2) & 37 & 1249 \\
ESO Slice Project (ESP) & 24 & 939 \\
The 6dF Galaxy Survey (6dFGS) & 22 & 535 \\ \hline
Baryonic Acoustic Oscillations Survey (BOSS) & 20 & 370 \\
Galaxy And Mass Assembly (GAMA) & 19 & 213 \\
WiggleZ & 14 & 194 \\ \hline
  \end{tabular}
  }
 \end{center}
\vspace{1mm}

\noindent Note these numbers were determined on $20^{th}$ Dec 2011
using the SOA/NASA ADS Astronomy Query Form by searching for refereed
papers which contained abstract keywords based on the following
boolean logic: (galaxy or galaxies) and ($<$long survey name$>$ or
$<$short survey name$>$). The table is purely indicative and not
weighted by survey age or effective cost. My apologies in advance for
the many surveys not included.
\end{table}

\section{Selected Highlights}
It is obviously an impossible task to try and summarise all the papers
covered by the surveys shown in Table~2 but below I provide some
personal reflections on the areas which have most interested me
(apologies in advance for the obvious bias):

\subsection{Luminosity Functions}
An original motivation for many surveys has been the measurement of
the galaxy luminosity function which describes the space density of
galaxies. Over the past 10 years this has now been measured in
UV/optical and near-IR bands and from redshift zero to relatively
high-redshifts (see for example: Cole et al.,~2001; Bell et al.,~2003;
Blanton et al.,~2003, 2005; Bouwens et al.,~2006, 2007; Faber et
al.,~2007; Hill et al.,~2010; Robotham \& Driver 2011). Although the
bright-end is generally well defined and well behaved the faint-end
and the implied space-density of dwarf systems remains elusive. This
is because of both the Eddington bias as well as the intrinsic low
surface brightness nature of these systems which places their peak
central surface brightnesses below the detection thresholds of the
imaging surveys. However it is also becoming clear that while the
Universe is not filled with giant low surface brightness galaxies a
key issue is our ability to accurately recover the fluxes of even the
most luminous systems. Two key questions are: how much light are we
missing around luminous systems, and how many dwarfs lie below our
thresholds. The upcoming deep optical surveys (VST, PanSTARRs, DES,
LSST) should be able to clarify both issues.

\subsection{Stellar Mass Measurements and the Galaxy Stellar Mass Function}
The past ten years has also seen a movement away from luminosity
functions to the more fundamental stellar mass functions. Motivated by
credible stellar mass estimates (e.g., Bell \& de Jong 2001; Kauffmann
et al.,~2003a; Bundy et al.,~2006; Taylor et al.,2011). Typically
alternative mass estimates agree to within a factor of 2 which has
allowed the construction of relatively consistent, and therefore
presumably robust, stellar mass functions (e.g., Bell et al.,~2003;
Baldry et al.,~2008, 2011; Fontana et al.,~2004; Ilbert et
al.,~2010). Our Universe appears to have a stellar mass density of
order $\Omega_* = 0.0017$ (4\% of the baryonic mass density, Baldry et
al.,~2011) with the evolution of the mass functions implying a
relatively linear build-up of log stellar mass with redshift (see
Sawicki~2011). One key issue is that the current estimates of stellar
mass appear to be a factor of two higher than what one expects from
the cosmic star-formation history using a standard IMF (Wilkins,
Trentham \& Hopkins 2008). This issue had also been explored earlier
by Baldry \& Glazebrook (2003) who proposed a modified-IMF to resolve
this issue which opens Pandora's Box on the question of an evolving
IMF (see for example Dave 2008).

\subsection{Galaxy Bimodality - Bimodality or Duality?}
Perhaps one of the strongest themes has been the rediscovery
(c.f. Baum 1959) of galaxy bimodality (e.g., Strateva et al.,~2001;
Baldry et al.,~2004, 2006; Driver et al.,~2006). Samples are now
routinely divided into red and blue subsets. Like Christmas LEDs which
spontaneously change colour as star-formation is quenched, unquenched
and quenched again. Looking at the images in Fig.~4 I'm tempted to
think the key point has been missed. On the whole the red and blue
samples show not only a marked difference in colour but also in
morphology. While quenching and unquenching may be a plausible
explanation for the colour change it is significantly harder to
identify a mechanism which readily modifies, compresses, and stretches
the morphologies (defined by the energy in the stellar orbits). In my
mind it seems the more elegant way forward is to recognise not the
{\it bimodality} of galaxies but the {\it duality}, or dual nature,
with galaxies typically comprising of a bulge and/or disc component.
In looking at Fig.~4 it is also clear that the red sample is a mixed
bag containing: spheroids, anemic spirals, and reddened spirals. This
becomes more apparent when one divides the sample by S\'ersic index
rather than colour and finds an equally distinct bimodality but with
less overlap with the colour-split sample than one would like.

Unfortunately considering the dual nature of galaxies is far harder
than a simple global bimodality split as it requires bulge-disc
decomposition which in turn requires high spatially resolved high
signal-to-noise data. Numerous groups are now embarked on this
endeavour both at low and high redshift (Kelvin et al.,~2011; Simard
et al.,~2011). One surprise which did stem from early work in this
direction (Allen et al.,~2006; Driver et al.,~2007a) is that while the
red peak might well contain 60\% of the stellar mass only half of this
is in spheroid structures such that only 40\% of the stellar mass in
total resides in spheroids and the remaining 60\% in discs. This is
perhaps at odds with a CDM hierarchical merger picture but, confirmed
by Gadotti (2009) and Tasca \& White (2011), appears to be a result
that is here to stay. So how then in a backdrop of hierarchical
merging, in which discs are easily disrupted, can the majority of the
stellar mass reside in dynamically thin fragile discs? It is also
worth noting that this must be a lower limit as numerous discs must
form which ultimately merge into spheroids such that some component of
the spheroids stellar mass also formed via the disc formation
process. In fact it begs the questions to whether merger-induced
star-formation is an almost negligible phenomena and that the majority
of stars are actually formed via direct gas infall -- as now argued by
some simulations (L'Huillier, Combes \& Semelin~2011) and directly
observed (e.g., Sancisi et al.,~2008).

\begin{figure}[h]

\vspace{-1.5cm}

\includegraphics[width=4.0in,angle=90]{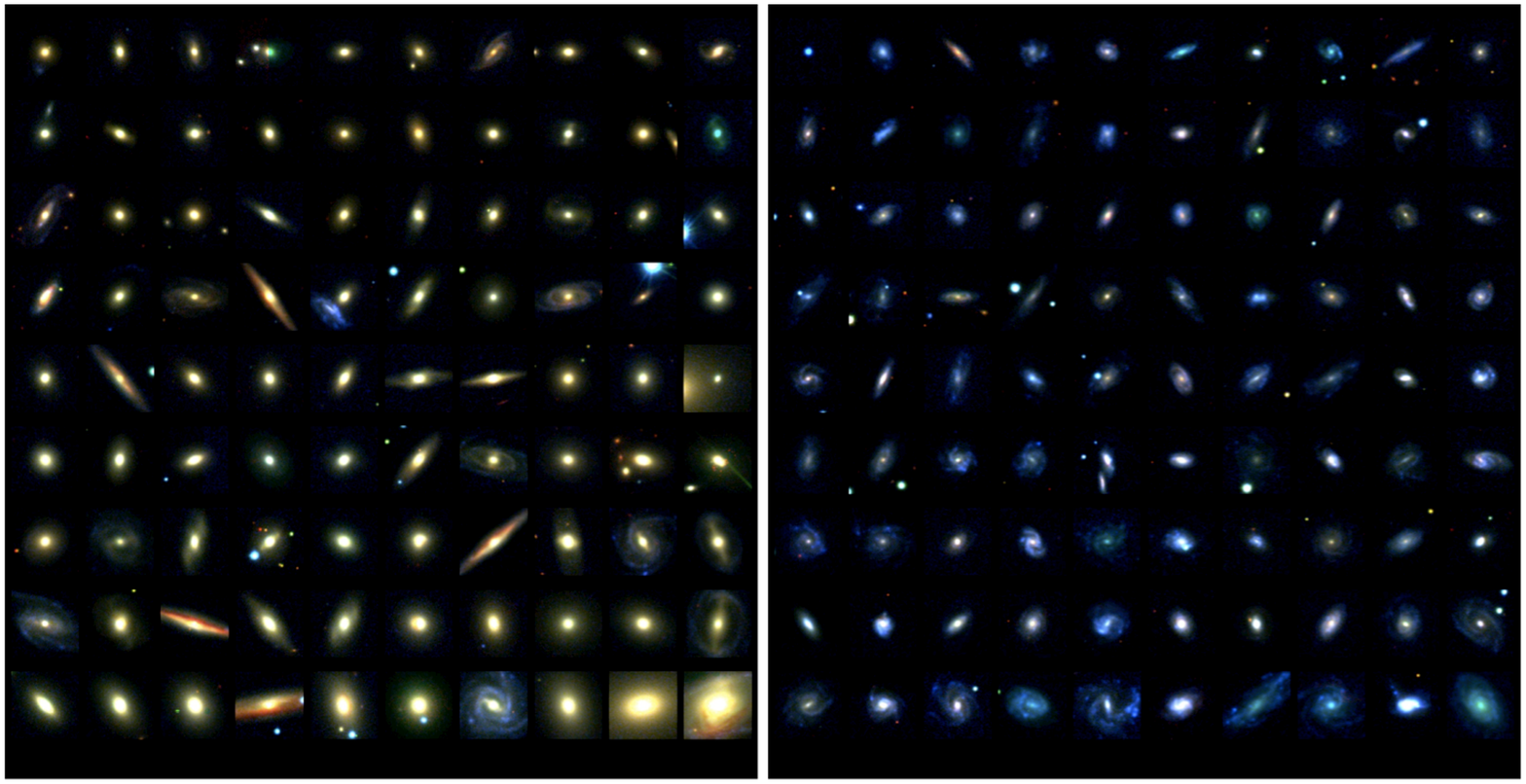} 

\vspace{-1.0cm}

 \caption{A sample of red (left) and blue (right) galaxies}

\end{figure}

\subsection{Dust Attenuation}
As highlighted by the obvious very red edge-on spiral interlopers in
Fig.~4 we see than dust attenuation is a significant factor typically
contaminating red samples at the 5-10\% level (simply from counting
the obvious edge-on systems in Fig.~4). While we now have a good
understanding of the dust attenuation law (e.g., Calzetti et
al.,~2000), and viable radiative transfer models (Popescu et
al.~2011), what is not clear is how the dust properties vary with
galaxy type, environment, or redshift. Although somewhat low profile
work Choi, Park \& Vogeley (2007), Shao et al., 2007, Driver et al
(2007b; 2008) and Masters et al.,~(2010) have all highlighted the
severe impact that dust attenuation can have with implied optical
bandpass corrections measured in magnitudes. In Driver et al.,~(2008)
we showed that in the $B$ band discs may be attenuated by as much as 1
magnitude while bulges by upto 2-3 magnitudes --- depending on
inclination. This is a result which is shocking to the optical
astronomers who routinely ignore intrinsic dust attenuation, but well
established to the far-IR community. Surveys such as GAMA (Driver et
al.,~2011) which aim to combine UV, optical, near-IR, mid-IR and
far-IR with a high fidelity spectroscopic data will be vital to
untangle this mess.

\subsection{The mass-metalicity relation}
A particularly elegant result of the past ten years has been the
beautiful mass-metalicity relation identified in SDSS and other
datasets (see in particular: Tremonti et al.,~2004; Savaglio et
al.,~2005; Erb et al.,~2006). Baldry, Glazebrook \& Driver (2008)
argue that this relation is an inevitable consequence of the variation
in star-formation efficiency with stellar mass. It could also be a
combination of star-formation efficiency (i.e., fraction of gas used,
inflows/outflows and IMF variation). A possible complicating factor
comes from consideration of the IGIMF (Wiedner \& Kroupa 2006) in
which more massive galaxies preferentially host more massive
star-formation regions leading to more enriched gas because the upper
end of the IMF is more likely to be populated

\subsection{The growth of galaxy sizes}
Slightly lower profile but entirely orthogonal is the work on
mass-size relations (e.g., Shen et al.,~2003; Blanton et al.,~2005;
Driver et al.,~2005). This has the potential to connect observable
galaxy sizes with the underlying halo spin parameter (i.e., the old
Fall \& Efstathiou 1988 connection, see also Dalcanton, Spergel \&
Summers 1998 and Mo, Mao and White 1998). More effort needs to be
invested in the numerical and simulation side however very nice clean
cut empirical results seem to be emerging the work of Trujillo et
al.~(2006; 2007) which finds significant evolution in galaxy sizes
with redshift. The obvious implication is that galaxies are visibly
growing from the inside out. In my world of duality this fits rather
nicely with the idea that we're witnessing the late formation and
growth of discs around the old pre-formed bulges. However other ideas
which involve a dynamical relaxation in which the galaxy core moves
inward while the outer regions move outwards have also been proposed.

\subsection{The cosmic star-formation rate}
Finally one cannot not fail to mention the great advances in measuring
star-formation rates starting from the seminal work of the CFRS and
HDF studies (Lilly et al.,~1996; Madau et al.,~1996), coupled with the
more local and exhaustive studies by Kauffmann et al.,~(2003b);
Brinchmann et al.,~(2004); Juneau et al.,~(2005) Salim et
al.,~(2007). The compilation of Hopkins \& Beacom (2006) provides what
appears to be a very firm and mature insight into the overall cosmic
star-formation history history with subsequent studies now finding
significant distinction by mass (e.g., Pozzetti et al.~2007) --- i.e.,
cosmic downsizing.

\section{Summary}
The above is a very brief pr\'ecis which tries in some way to
acknowledge the efforts of those who have designed and led these
surveys, to temper our appetite for photometric redshifts with some
words of caution, and to mention in passing a small fraction of the
high-note papers and results in this subject over the past decade. The
next decade will see survey expansion continue as we push deeper at
optical wavelengths (VST, VISTA, PanSTARRs, DES, LSST, Euclid), but
perhaps more significant the expansion in wavelength with public
releases imminent of the WISE mid-IR and Herschel Far-IR data. These
in turn will be followed by major X-ray (XMM-XLL) and Radio programs
(ASKAP, MeerKAT) which will enable the start of truly panchromatic
surveys which will allow for the comprehensive and simultaneous study
of the AGN, gas, dust, and stellar components as a function of
environment, and distance. These mega-surveys will then feed Integral
Field Unit (IFU) and multi-IFU follow-up which will provide spatial
studies of the internal dynamics and chemistry of well defined
sub-samples. Exciting times.

I'd like to thank the organisers for a very enjoyable meeting and
conclude by dedicating this article to the memory of John Huchra who
really started something.

\pagebreak

\tiny

\begin{description}
\item Allen P.D., Driver S.P., Graham A.W., Cameron E., Liske J., De Propris R., 2006, MNRAS, 371, 2
\item Baldry I.K., Glazebrook K., 2003, ApJ, 593, 258
\item Baldry I.K., Glazebrook K., Brinkmann J., Ivezic Z., Lupton R.H., Nichol R.C., Szalay A.S., 2004, ApJ, 600, 681
\item Baldry I.K., et al., 2005, MNRAS, 358, 441
\item Baldry I.K., Balogh M.L., Bower R.G., Glazebrook K., Nichol R.C., Bamford S.P., Budavari T., 2006, MNRAS, 373, 469
\item Baldry I.K., GLazebrook K., Driver S.P., 2008, MNRAS, 388, 945
\item Baldry I.K., et al., 2010, MNRAS, 404, 86
\item Baldry I.K., et al., 2011, MNRAS, in press (arXiv:1111.5707)
\item Baum W.A., 1959, PASP, 71, 106 
\item Bell E.F., de Jong R.S., 2001, ApJ, 550, 212
\item Bell E.F., McIntosh D.H., Katz N., Weinberg M.D., 2003, ApJS, 149, 289
\item Blanton M.R., et al., 2003, ApJ, 592, 819
\item Blanton M.R., et al., 2005, AJ, 129, 2562
\item Bouwens R.J., Illingworth G.D., Blakeslee J.P., Franx M., 2006, ApJ, 653, 53
\item Bouwens R.J., Illingworth G.D., Franx M., Ford H., 2007, ApJ, 670, 928
\item Brinchmann J., Charlot S., White S.D.M., Tremonti C., Kauffmann G., Heckman T., Brinkmann J., 2004, MNRAS, 351, 1151
\item Bundy K., et al., 2006, ApJ, 651, 120
\item Calzetti D., Armus L., Bohlin R.C., Kinney A.L., Koornneef J., Storchi-Bergmann T., 2000, ApJ, 533, 682
\item Choi Y-Y., Park C., Vogeley M.S., 2007, ApJ, 658, 884
\item Cole S., et al., 2001, MNRAS, 326, 255
\item Collister A., Lahav O., 2004, PASP, 116, 345
\item Canon R., et al., 2006, MNRAS, 372, 425
\item Colless M.M., et al., 2001, MNRAS, 328, 1039
\item da Costa L., et al., 1998, AJ, 116, 1
\item Dave R., 2008, 385, 147<
\item Davis M., Huchra J., Latham D.W., Tonry J., 1982, ApJ, 253, 423
\item Davis M., et al., 2003, SPIE, 4843, 161
\item Dalcanton J., Spergel D.N., Summers F.J., 1997, ApJ, 482, 659
\item Drinkwater M.J., et al., 2010, MNRAS, 401, 1429
\item Driver S.P., Liske J., Cross N.J.G., De Propris R., Allen P.D., 2005, MNRAS, 360, 81
\item Driver S.P., Allen P.D., Graham A.W., Cameron E., Liske J., Ellis S.C., Cross N.J.G., De Propris R., Phillipps S., Couch W.J., 2006, MNRAS, 368, 414
\item Driver S.P., Allen P.D., Liske J., Graham A.W., 2007, ApJ, 657, 85
\item Driver S.P., Popescu C.C., Tuffs R.J., Liske J., Graham A.W., Allen P.D., De Propris R., 2007, MNRAS, 379, 1022
\item Driver S.P., Popescu C.C., Tuffs R.J., Graham A.W., Liske J., Baldry I.K., 2008, ApJ, 678, 101
\item Driver S.P., et al., 2011, MNRAS, 413, 971
\item Eisenstein D.J., et al., 2001, AJ, 122, 2267
\item Ellis R.S., Colless M., Broadhurst T., Heyl J., Glazebrook K., 1996, MNRAS, 280, 235
\item Erb D.S., Shapley A.E., Pettini M., Steidel C.C., Reddy N.A., Adelberger K.L., 2006, ApJ, 644, 813
\item Faber S., et al., 2007, ApJ, 665, 265
\item Falco E.E., et al., 1999, PASP, 758, 438
\item Fall S.M., Efstathiou G., 1980, MNRAS, 193, 189
\item Fontana A., et al., 2004, A\&A, 424, 23
\item Gadotti D., 2009, MNRAS, 393, 1531
\item Garilli B., et al., 2008, A\&A, 486, 683
\item Geller M., et al., 1997, AJ, 114, 2205
\item Giavalisco M., et al., 2004, ApJ, 600, L93
\item Gunawardhana M.L.P., et al., 2011, MNRAS, 415, 1647
\item Haas M.R., Schaye J., Jeeson-Daniel A., 2011, MNRAS, in press (arXiv: 1103.0547) 
\item Hopkins A.M., Beacom J.F., 2006, ApJ, 651, 142
\item Hill D.T., Driver S.P., Cameron E., Cross N.J.G., Liske J., Robotham A., 2010, MNRAS, 404, 1215
\item Huang J.-S., Glazebrook K., Cowie L.L., Tinney C., 2003, ApJ, 584, 203
\item Ilbert O., et al., 2010, ApJ, 709, 644
\item Jones H., et al., 2009, MNRAS, 399, 683 
\item Juneau S., et al., 2005, ApJ, 619, 135
\item Kauffmann G., et al., 2003a, MNRAS, 341, 33
\item Kauffmann G., et al., 2003b, MNRAS, 341, 54
\item Kelvin L.S., et al., 2012, MNRAS, in press (arXiv:1112.1956)
\item Kochaneck C.S., et al., 2011, ApJS, in press (arXiv:1110.4371)
\item L'Huillier B., Combes F., Semelin B., 2011, A\&A, in press (arXiv:1108.4247)
\item Le F\`evre O., et al., 2005, A\&A, 439, 845
\item Lilly S.J., Fevre O., Crampton D., Hammer F., Tresse L., 1995, ApJ, 455, 50
\item Lilly S.J., Le Fevre O., Hammer F., Crampton D., 1996, ApJ, 460, 1
\item Lilly S.J., et al., 2007, ApJS, 172, 70
\item Loveday J., Peterson B.A., Efstathiou G., Maddox S.J., 1992, ApJ, 390, 338
\item Madau P., Ferguson H.C., Dickinson M.E., Giavalisco M., Steidel C.C., Fruchter A., 1996, MNRAS, 283, 1388
\item Masters K., et al., 2010, MNRAS< 404, 792
\item Mo H.J., Mao S., White S.D.M., 1998, MNRAS, 295, 319
\item Peterson B.A., Ellis R.S., Efstathiou G., Shanks T., Bean A.J., Fong R., Zen-Long Z., 1986, MNRAS, 221, 233
\item Popescu C.C., Tuffs R.J., Dopita M.A., Fischera J., Kylafis N.D., Madore B.F., 2011, A\&A, 527, 109
\item Pozzetti L., et al., 2007, A\&A, 474, 443
\item Ratcliffe A., Shanks T., Broadbent A., Parker Q.A., Watson F.G., Oates A.P., Fong R., Collins C.A., 1996, MNRAS, 281, 47
\item Robotham A.S.G., Driver S.P., 2011, MNRAS, 413, 2570
\item Salim S., et al., 2007, ApJS, 173, 267
\item Sancisi R., Fraternall F., Oosterloo T., van der Hulst J.M., 2008, A\&A Rv, 15, 189
\item Sawicki M., 2011, MNRAS, in press (arXiv:1108.5186)
\item Savaglio S., et al., 2005, ApJ, 635, 260
\item Schectman S.A., Landy S.D., Oemler A., Tucker D.L., Lin H., Kirschner R.P., Schechter P.L., 1996, ApJ, 470, 172
\item Shao Z., Xiao Q., Shen S., Mo H.J., Xia X., Deng Z., 2007, ApJ, 659, 1159
\item Shen S., et al., 2003, MNRAS, 343, 978
\item Simard L., Mendel J.T., Patton D.R., Ellison S.L., McConnachie Alan W., 2011, ApJS, 196, 11
\item Strateva I., et al., 2001, AJ, 112, 1861
\item Strauss M.A., et al., 2002, AJ, 124, 1810
\item Tasca L.A.M., White S.D.M., 2011, A\&A, 530, 106
\item Taylor E., et al., 2011, MNRAS, 418, 1587
\item Tremonti C.A., et al., 2004, ApJ, 613, 898
\item Trujillo I., et al., 2006, ApJ, 650, 18
\item Trujillo I., et al., 2007, MNRAS, 382, 109
\item Vettolani G., et al., 1997, 325, 954
\item Wilkins S.M., Trentham N., Hopkins A., 2008, MNRAS, 385, 687
\item Wolf C., Meisenheimer K., Rix H.-W., Borch A., Dye S., Kleinheinrich M., 2003, A\&A, 401, 73
\item Wolf C., Hildebrandt H., Taylor E.N., Meisenheimer K., 2008, A\&A, 492, 933 
\item Yee H.K.C., et al., 2000, ApJS, 129, 475
\item York D., et al., 2000, AJ, 120, 1579
\end{description}

\end{document}